\documentclass[proceedings]{rmaa}

\usepackage{rmaacite}

\usepackage{flushrt}
\usepackage[figuresright]{rotating}

\renewcommand{\P}[1]{%
\ifnum#1=1\hbox{OW~168--326E}\fi
\ifnum#1=2\hbox{OW~167--317}\fi
\ifnum#1=3\hbox{OW~163--317}\fi
\ifnum#1=5\hbox{OW~158--323}\fi
\ifnum#1=0\hbox{OW~171--334}\fi}

\title{Reionization of the Universe 
and the Photoevaporation of Cosmological Minihalos}
\author{Paul R. Shapiro
  \affil{University of Texas at Austin, USA}
Alejandro C. Raga
 \affil{Instituto de Astronom\'\i a,
  UNAM, M\'exico}}

\fulladdresses{
\item Paul R. Shapiro,
  Dept. of Astronomy, University of Texas, Austin, TX 78712
  (shapiro@astro.as.utexas.edu)
\item Alejandro C. Raga, Instituto de Astronom\'\i a,
  UNAM, Apartado Postal 70-264, 04510 M\'exico D. F.,
M\'exico.}

\shortauthor{Shapiro \& Raga}
\shorttitle{Reionization and Minihalo Photoevaporation}


\abstract{
The first sources of ionizing radiation to condense out of the dark and
neutral IGM sent ionization fronts sweeping outward through their
surroundings, overtaking other condensed objects and photoevaporating them.
This feedback of universal reionization on cosmic structure 
formation is demonstrated here by gas dynamical simulations,
including radiative transfer, for a cosmological minihalo of
dark matter and baryons exposed to an external source of ionizing radiation,
 either starlight or quasar light,
just after the passage of the global 
ionization front created by the source.}

\listofauthors{P.~R.~Shapiro \& A.~C.~Raga}
\indexauthor{Shapiro, P.~R.}
\indexauthor{Raga, A.~C.}

\begin{document}

\maketitle


\section{Ionization Fronts in the IGM} 

The neutral, opaque IGM
out of which the first bound objects condensed was dramatically reheated
and reionized at some time between a redshift $z\approx50$ and $z\approx5$
by the radiation released by some of these objects.
When the first sources turned on, they
ionized their surroundings by propagating weak, R-type
ionization fronts which moved outward supersonically with respect to both
the neutral gas ahead of and the ionized gas behind the front, racing ahead
of the hydrodynamical response of the IGM, as first described by 
Shapiro (1986) and Shapiro \& Giroux (1987). These authors solved the
problem of the time-varying radius of a spherical I-front which surrounds 
isolated sources in a cosmologically-expanding IGM analytically, taking
proper account of the I-front jump condition generalized to cosmological
conditions. They applied these solutions to determine when the I-fronts
surrounding isolated sources would grow to overlap and, thereby,
complete the reionization of the universe.
The effect of density inhomogeneity on the rate of I-front propagation
was described by a mean ``clumping factor'' $c_l>1$, which
slowed the I-fronts by increasing the average recombination rate per H atom
inside clumps. This suffices to describe the rate of I-front propagation
as long as the
clumps are either not self-shielding or, if so, only
absorb a fraction of the ionizing photons emitted by the central source. 
Numerical radiative transfer methods are currently under
development to solve this problem in 3D for the inhomogeneous density 
distribution which arises as cosmic structure forms, so far limited to a 
fixed density field without gas dynamics (e.g.\ Abel, Norman, \& Madau 1999;
Razoumov \& Scott 1999; Ciardi et al. 2000). A different, more approximate approach which
is intended to mimic the average rate at which I-fronts expanded and 
overlapped during reionization within the context of cosmological gas 
dynamics simulation has also been developed (Gnedin 2000).
The question of what
dynamical effect the I-front had on the density inhomogeneities it
encountered, however, requires further analysis. Here we shall briefly summarize
our results of the first radiation-hydrodynamical simulations of the back-reaction
of a cosmological I-front on a gravitationally-bound density inhomogeneity
it encounters -- a dwarf galaxy minihalo -- during reionization.  

\section{The Photoevaporation of Dwarf Galaxy Minihalos
Overtaken by a Cosmological Ionization Front}

\begin{figure}
\vspace{-0.7in}
\begin{minipage}[c]{.55\textwidth}
\centering
\includegraphics[width=0.85\textwidth]{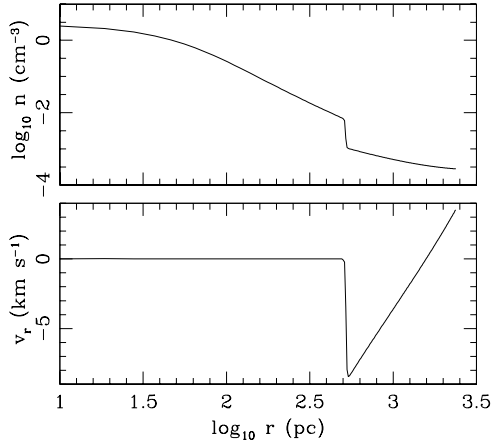}
\end{minipage}
\begin{minipage}[c]{.35\textwidth}
\caption{MINIHALO INITIAL CONDITIONS BEFORE REIONIZATION.
(Top) gas density; 
(Bottom) gas velocity versus distance from minihalo  center.}
\end{minipage}
\vspace{-0.8in}
\end{figure}

We have performed radiation-hydrodynamical simulations
of the photoevaporation of a cosmological minihalo overrun by a weak,
R-type I-front in the surrounding IGM, created by an external source 
of ionizing radiation (Shapiro and Raga 2000a,b).
Our simulations in 2D, axisymmetry used an
Eulerian hydro code with Adaptive Mesh Refinement and
the Van~Leer flux-splitting algorithm, which
solved nonequilibrium ionization rate equations (for H, He, C, N, O, Ne,
and S) and included an explicit treatment of radiative transfer
by taking into account the bound-free opacity of H and He. 
A possible heavy element abundance of $10^{-3}$ times solar was assumed,
 as well. 

{  Here we compare some of those results for two different
sources: a quasar-like source with emission spectrum 
$  \it F_\nu\propto\nu^{-1.8}$ ($\nu>\nu_H$) and a stellar 
source with a 50,000 K blackbody spectrum}, with luminosity 
and distance adjusted to keep the ionizing photon fluxes the 
same in the two cases. In particular, if $r_{\rm Mpc}$ is the
distance (in Mpc) between source and minihalo and $N_{{\rm ph},56}$ is the 
H-ionizing photon luminosity (in units of $10^{56}\,s^{-1}$), then the
flux at the location of the minihalo would, if unattenuated,
correspond to $N_{{\rm ph},56}/r^2_{\rm Mpc}=1$.
\begin{figure}
\centering
\includegraphics[height=2.4in,width=3in]{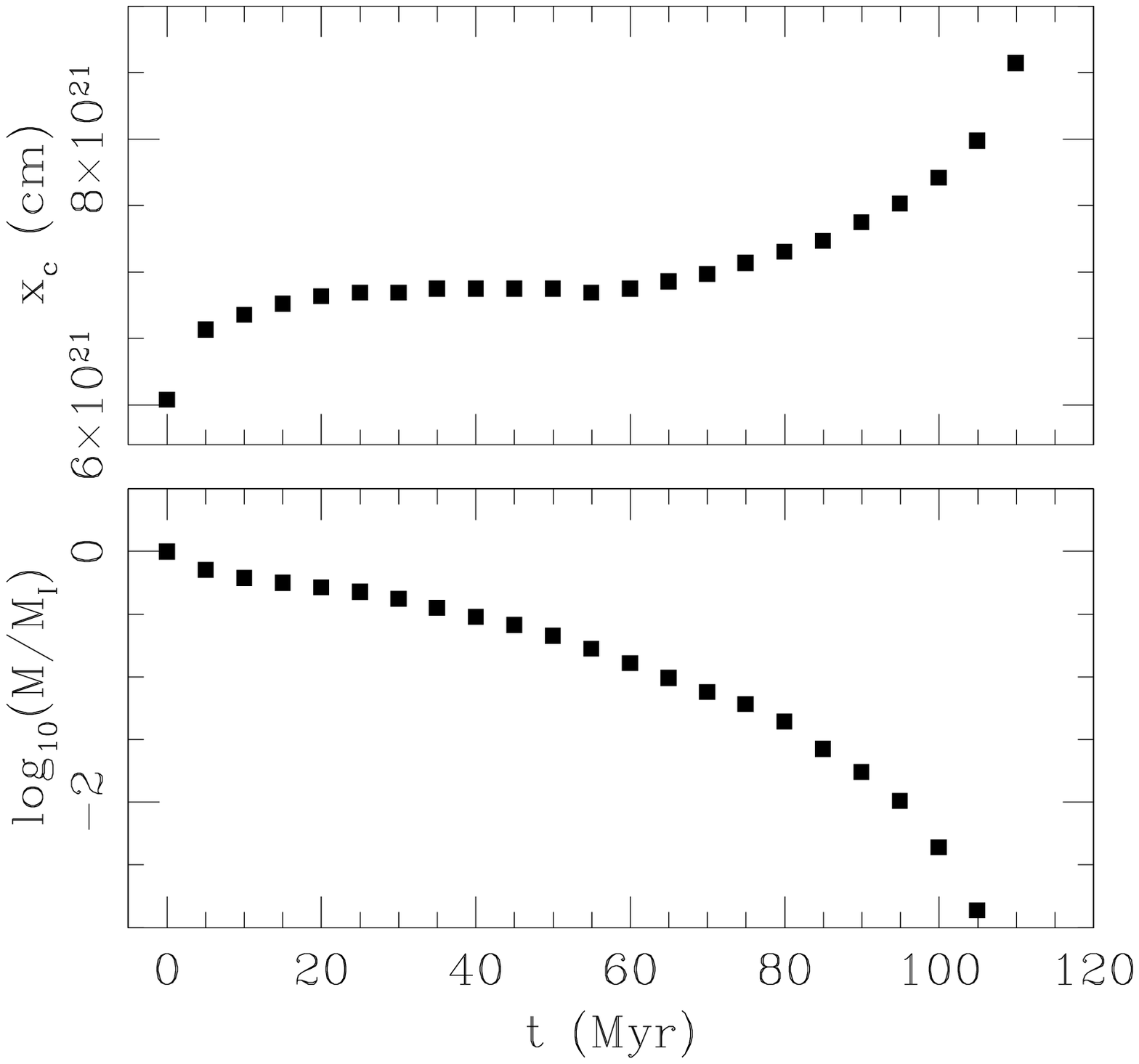}
\includegraphics[height=2.4in,width=3in]{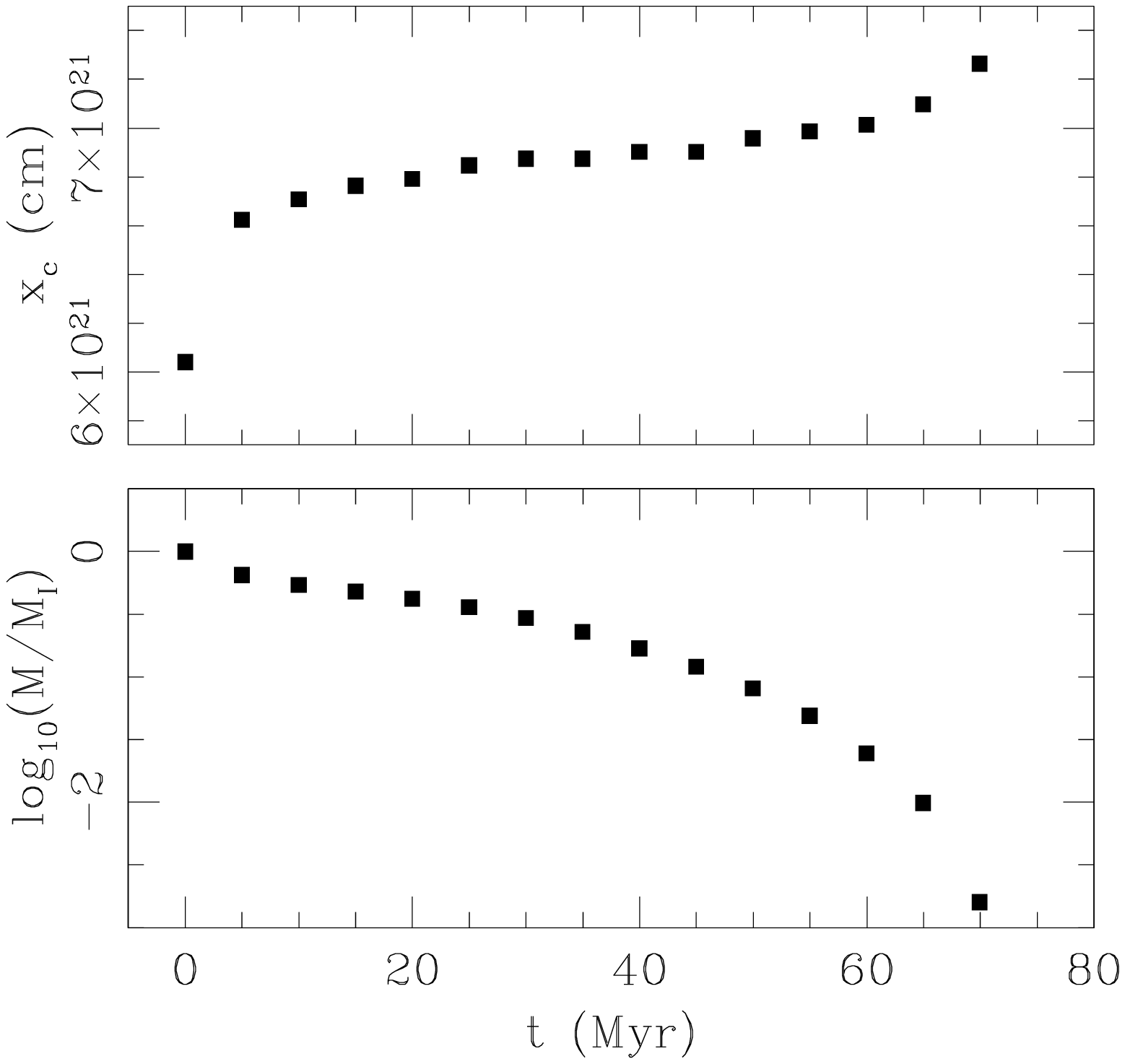}
\vspace{-0.5in}
\caption{I-FRONT PHOTOEVAPORATES MINIHALO: (upper panels)
 I-front position along $x$-axis versus time; (lower panels) Mass
fraction of the initial mass $M_{\rm I}$ of minihalo hydrostatic 
        core which remains neutral (H~I) versus time.
(a) (left) STELLAR CASE (b) (right) QUASAR CASE.}
\end{figure}
{  Our initial condition before ionization, shown in Figure 1, is
that of a $10^7M_\odot$ minihalo in an Einstein-de~Sitter
universe ($\Omega_{\rm CDM}=1-\Omega_{\rm bary}$; $\Omega_{\rm
bary}h^2=0.02$;
$h=0.7$) which collapses out and virializes at $z_{\rm coll}=9$,}
yielding a truncated, nonsingular isothermal sphere of radius 
$R_c=0.5\,\rm kpc$ in hydrostatic equilibrium
with virial temperature $T_{\rm vir}=5900\,\rm K$
and dark-matter velocity dispersion $\sigma_V=6.3\,\rm km\,s^{-1}$,
according to the solution of Shapiro, Iliev, \& Raga (1999),
for which the finite central density inside a radius about 1/30 of the
total size of the sphere 
is 514 times the surface density. This
hydrostatic core of radius $R_c$ is embedded in a self-similar,
spherical, cosmological infall according to Bertschinger (1985).

{  The results of our simulations on an $  (r,x)$-grid with 
$  256\times512$ cells (fully refined), summarized in Figures~2--6, include
the following points:}
\begin{itemize}
\item The background IGM and infalling gas outside the minihalo [centered at
$(r,x)=(0,7.125\times10^{21}\,{\rm cm})$] are quickly
ionized, and the resulting pressure gradient in the infall region
converts the infall into an outflow. 
\item As expected, the hydrostatic core of the minihalo
 shields itself against ionizing photons,
trapping the I-front which enters the halo, causing it to decelerate
inside the halo to the sound speed of the ionized gas before it can
exit the other side, thereby transforming itself into a weak, D-type
front preceded by a shock. 

\item The side facing the source
expels a supersonic wind backwards towards the source, which shocks
the IGM outside the minihalo, while the remaining neutral halo material
is accelerated away from the source by the so-called ``rocket effect''
as the halo photoevaporates (cf.\ Spitzer 1978). Since this is a case
of gas bound to a dark halo with 
$\sigma<\rm 10\,km\,s^{-1}$, this photoevaporation proceeds unimpeded by
gravity. 
\begin{figure}
\includegraphics[width=0.5\textwidth]{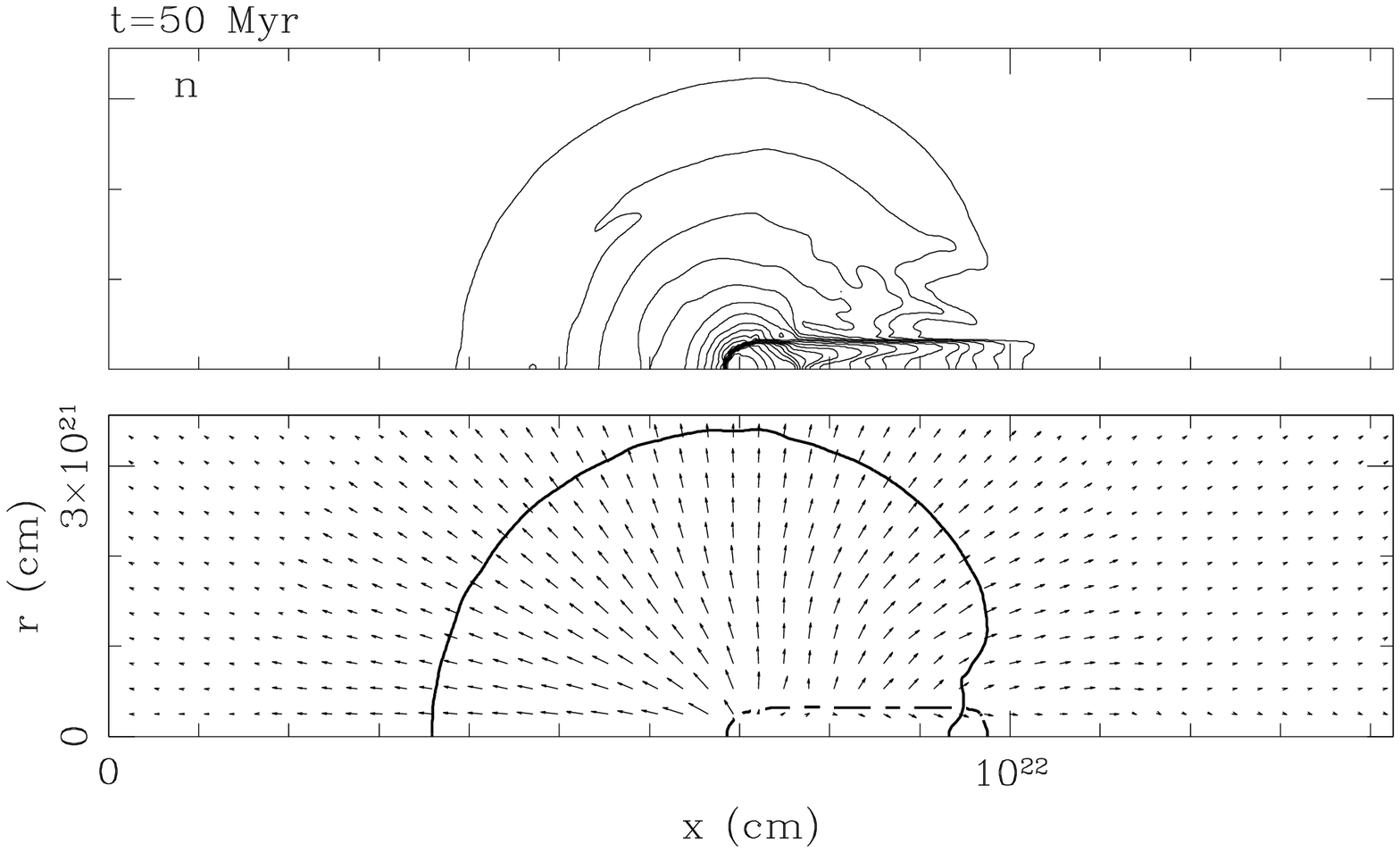}
\includegraphics[width=0.5\textwidth]{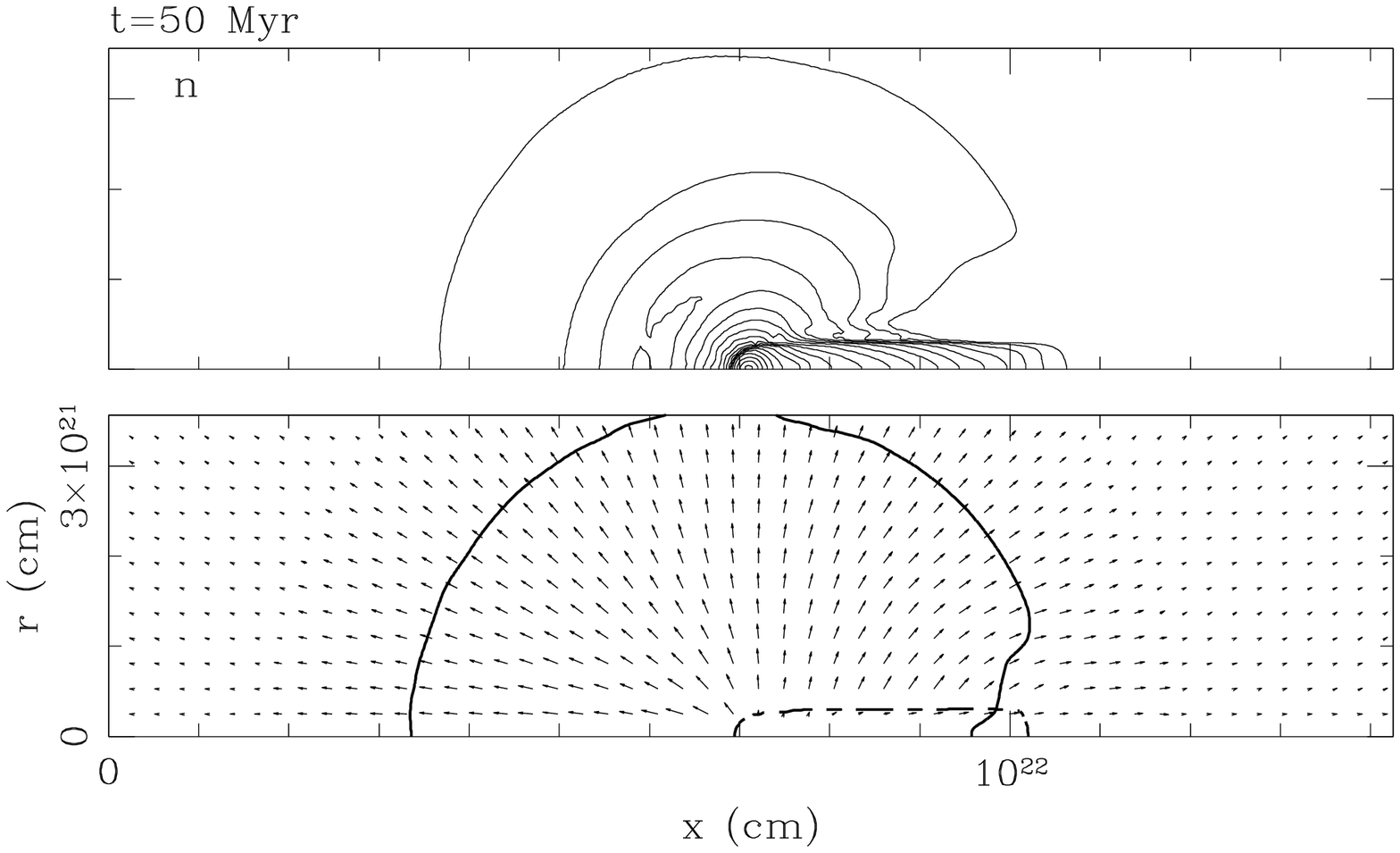}
\vspace{-1.3in}
\caption{PHOTOEVAPORATING MINIHALO:
One time-slice 50 Myr after turn-on of ionizing source to
 the left of computational box along the $x$-axis. 
(a) (left) STELLAR CASE. (upper panel) isocontours of atomic density, 
logarithmically spaced, 
in $(r,x)-$plane of cylindrical coordinates; (lower panel)
velocity arrows are plotted with length proportional to gas velocity.
An arrow of length equal to the spacing between arrows has velocity
$25\, {\rm km\, s^{-1}}$. Solid line shows current extent of gas originally in
hydrostatic core. Dashed line is I-front (50\% H-ionization contour).
(b) (right) QUASAR CASE.
Same as (a), except for quasar source, instead, and
an arrow of length equal to the spacing between arrows has velocity
$30\, {\rm km\, s^{-1}}$. }
\end{figure}
\item Figures~2(a) and (b) show the position of the I-front inside
the minihalo as it slows from weak, R-type to weak D-type led by a shock
as it advances across the original hydrostatic core, for the two cases. 
Figures~2(a) and (b) also show the mass of the
neutral zone within the original hydrostatic core shrinking as
the minihalo photoevaporates within about 100~Myrs. 
The photoevaporation time is 50~\% larger for the stellar source 
than for the quasar source. 

\item Figures~3 and 4
show the structure of the photoevaporative flow 50~Myrs after the
global I-front first overtakes the minihalo, with key features of
the flow indicated by the labels on the temperature plot in Figure~4.
For the stellar case, a strong shock labelled ``4S'' clearly leads the 
D-type I-front (labelled
``4I'') as it advances through the minihalo core, by contrast with the 
quasar case in which hard photons penetrate deeper into the neutral gas 
and preheat it, thereby weakening the shock which leads the I-front. 
The softer stellar spectrum also explains why helium on the ionized side
of the I-front is mostly He~II, rather than He~III as in the quasar case,
while the neutral side is completely He~I, rather than a mix of He~I and II
as in that case.    

\item Figure~5 shows the spatial variation of the relative abundances of
C, N, O ions along the symmetry axis after 50~Myrs.
While the quasar case shows the presence at 50~Myrs
of low as well as high ionization stages for the metals, the softer spectrum
of the stellar case yields less highly ionized gas on the ionized side of 
the I-front (e.g. mostly C~III, N~III, O~III) and the neutral side
as well (e.g. C~II, N~I, O~I and II).

\item The column densities of H~I, He~I and II, and C~IV for minihalo gas of
different velocities as seen along the symmetry axis at different times
are shown in Figure~6. 
At early times, the minihalo gas resembles a weak
Damped Lyman $\alpha$ (``DLA'') absorber with small velocity width 
($\geq 10\rm\,km\,s^{-1}$) and $N_{\rm H\,I}\geq 10^{20}\rm cm^{-2}$,
with a Lyman-$\alpha$-Forest(``LF'')-like red wing 
($\hbox{velocity width}\,\geq 10\,\rm km\,s^{-1}$)
with $N_{\rm H\,I}\geq10^{16}\rm cm^{-2}$ on the side moving toward
the source, with a He~I profile which mimics that of H~I but with 
$N_{\rm He\,I}/N_{\rm H\, I}\sim {\rm [He]/[H]}$, and with a weak C~IV 
feature with $N_{\rm C\,IV}\sim10^{11}\,(10^{12})\,\rm cm^{-2}$ for the
stellar (quasar) cases, respectively, displaced in
this same asymmetric way from the velocity of peak H~I column
density. For He~II at early times, the stellar case has 
$N_{\rm He\,II}\approx10^{18}\rm cm^{-2}$ shifted by 10's of 
$\rm km/sec$ to the red of the H~I peak, while for the quasar case,
He~II simply follows the H~I profile, except that 
$N_{\rm He\,II}/N_{\rm H\,I}\approx10$ in the red wing but 
$N_{\rm He\,II}/N_{\rm H\,I}\approx10^{-2}$ in the central H~I 
feature. After 160~Myr,
however, only a narrow H~I feature with LF-like column density
$N_{\rm H\,I}\sim10^{13}\,(10^{14})\rm cm^{-2}$ remains, with 
$N_{\rm He\,I}/N_{\rm H\, I}\sim 1/4\, (<1/10)$, 
$N_{\rm He\,II}/N_{\rm H\,I}\sim10^3\, (10^2)$, and 
$N_{\rm C\,IV}/N_{\rm H\,I}\sim3\rm[C]/[C]_\odot \,(\rm[C]/[C]_\odot)$ 
for the stellar (quasar) cases, respectively.

\item Observations of the absorption spectra of high redshift sources like those 
which reionized the universe may reveal the presence of photoevaporative 
flows like these and provide a useful diagnostic of the reionization process.
\item Future work will extend this study to minihalos
of higher virial temperatures, for which gravity competes more effectively
with photoevaporation.
\end{itemize}

\begin{figure}
\begin{minipage}[c]{4.5in}
\includegraphics[angle=90,height=4in,width=4.5in]{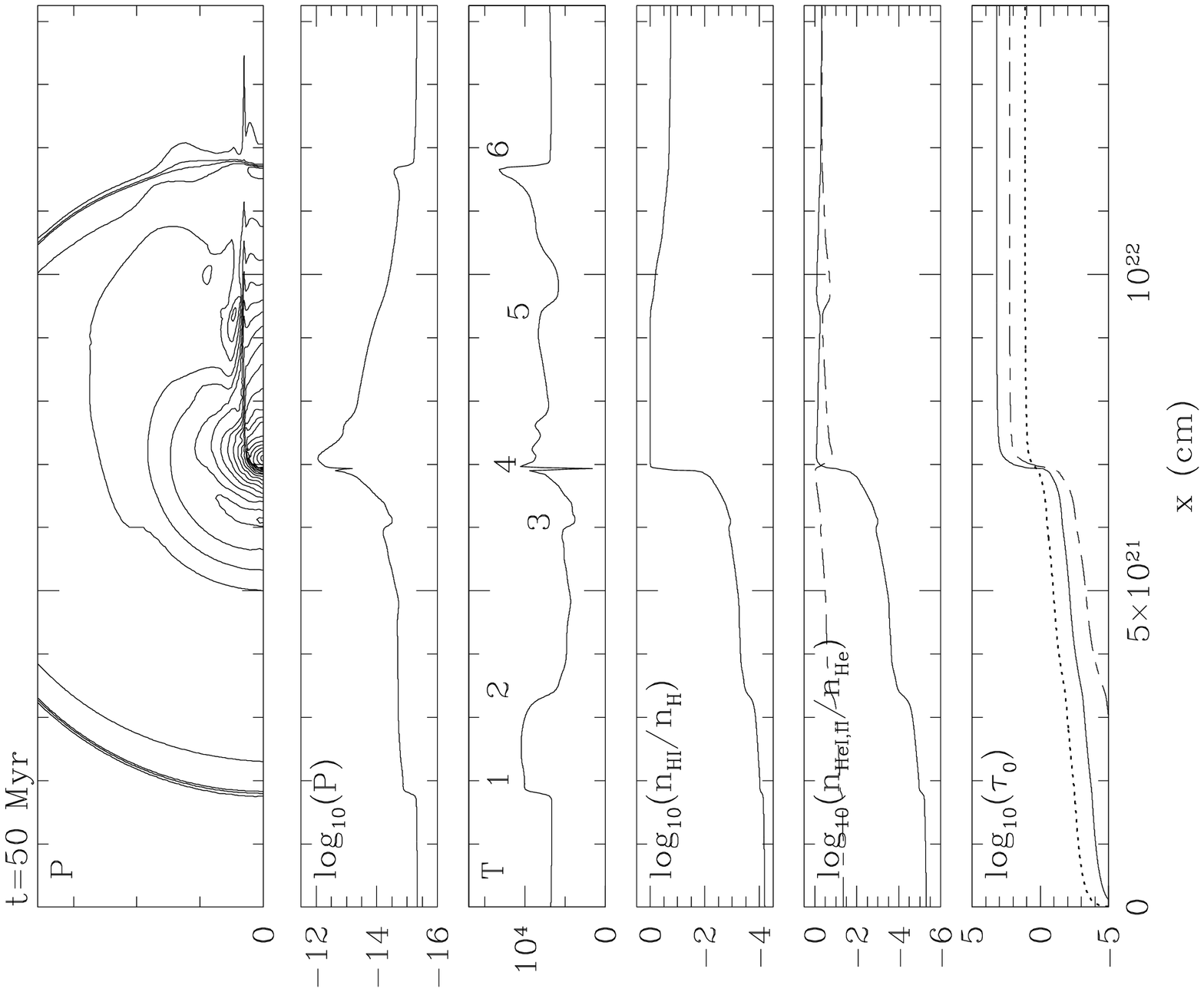}
\includegraphics[angle=90,height=4in,width=4.5in]{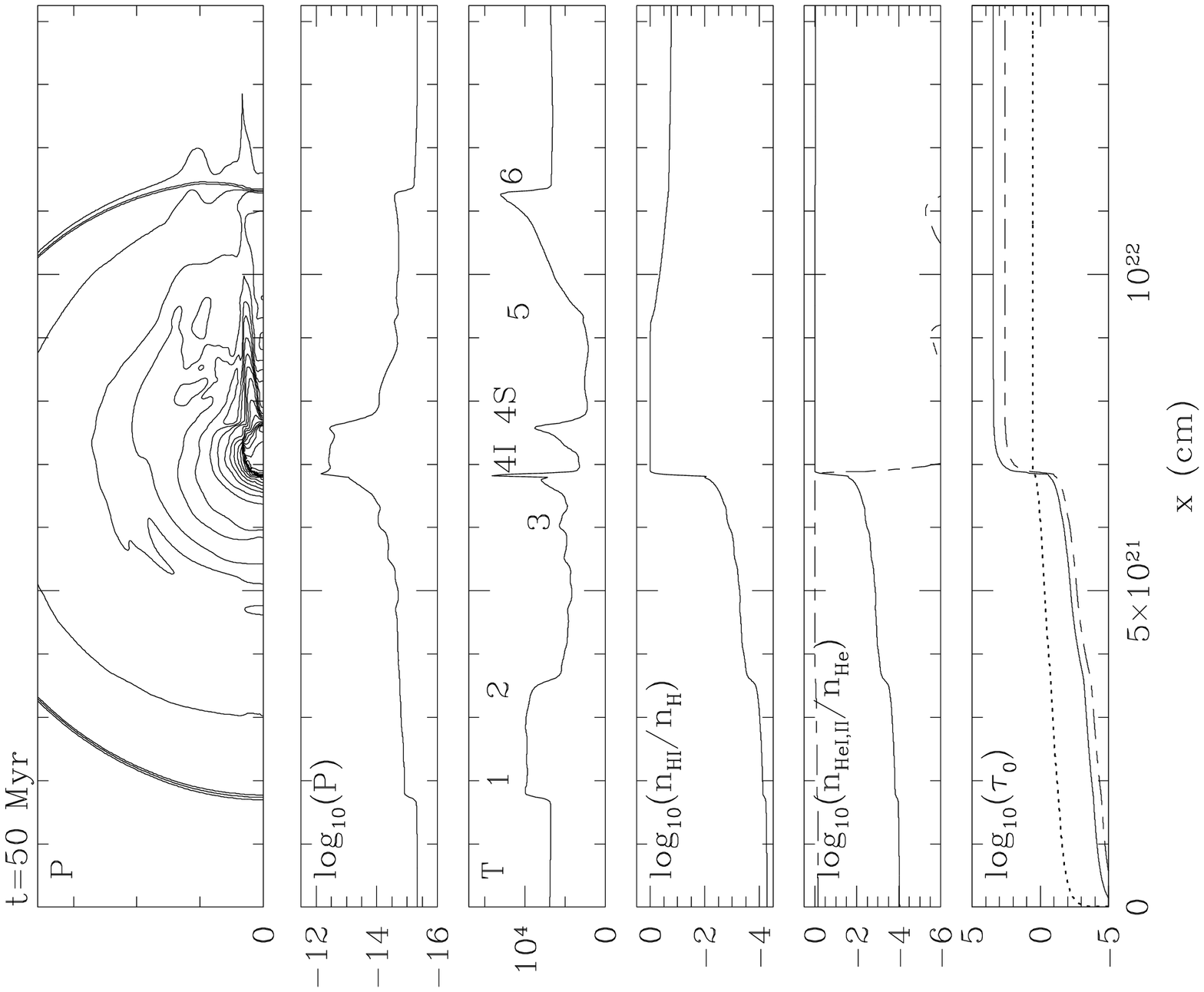}
\end{minipage}
\begin{minipage}[c]{6.5in}
\rotcaption{PHOTOEVAPORATING MINIHALO: 
	One time-slice 50 Myr after turn-on
      of ionizing source located to the left of
      computational box along the $x$-axis. (a) (left) STELLAR CASE. 
	From top to bottom: (i)
      isocontours of pressure, logarithmically spaced, in
      $(r,x)-$plane of cylindrical coordinates; (ii) pressure along the
      $r=0$ symmetry axis; (iii) temperature; (iv) H~I fraction; (v) He~I
      (solid) and He~II (dashed) fractions; (vi) bound-free optical
      depth along $r=0$ axis at the threshold ionization energies for
      H~I (solid), He~I (dashed), He~II (dotted). Key features of the
      flow are indicated by the numbers which label them on the
      temperature plot: 1~=~IGM shock; 2 = contact discontinuity
      between shocked halo wind and swept-up IGM; 3 = wind shock;
      between 3 and 4 = supersonic wind; 4I = I-front; 
      4S = shock which precedes
      I-front; 5 = boundary of
      gas originally in hydrostatic core; 6 = shock in shadow region
      caused by compression of shadow gas by shock-heated gas outside
      shadow. (b) (right) QUASAR CASE.
	Same as (a), except for quasar source, instead, and
	label 4 = I-front in temperature plot.}
\end{minipage}
\end{figure}
\begin{figure}
\includegraphics[width=0.5\textwidth]{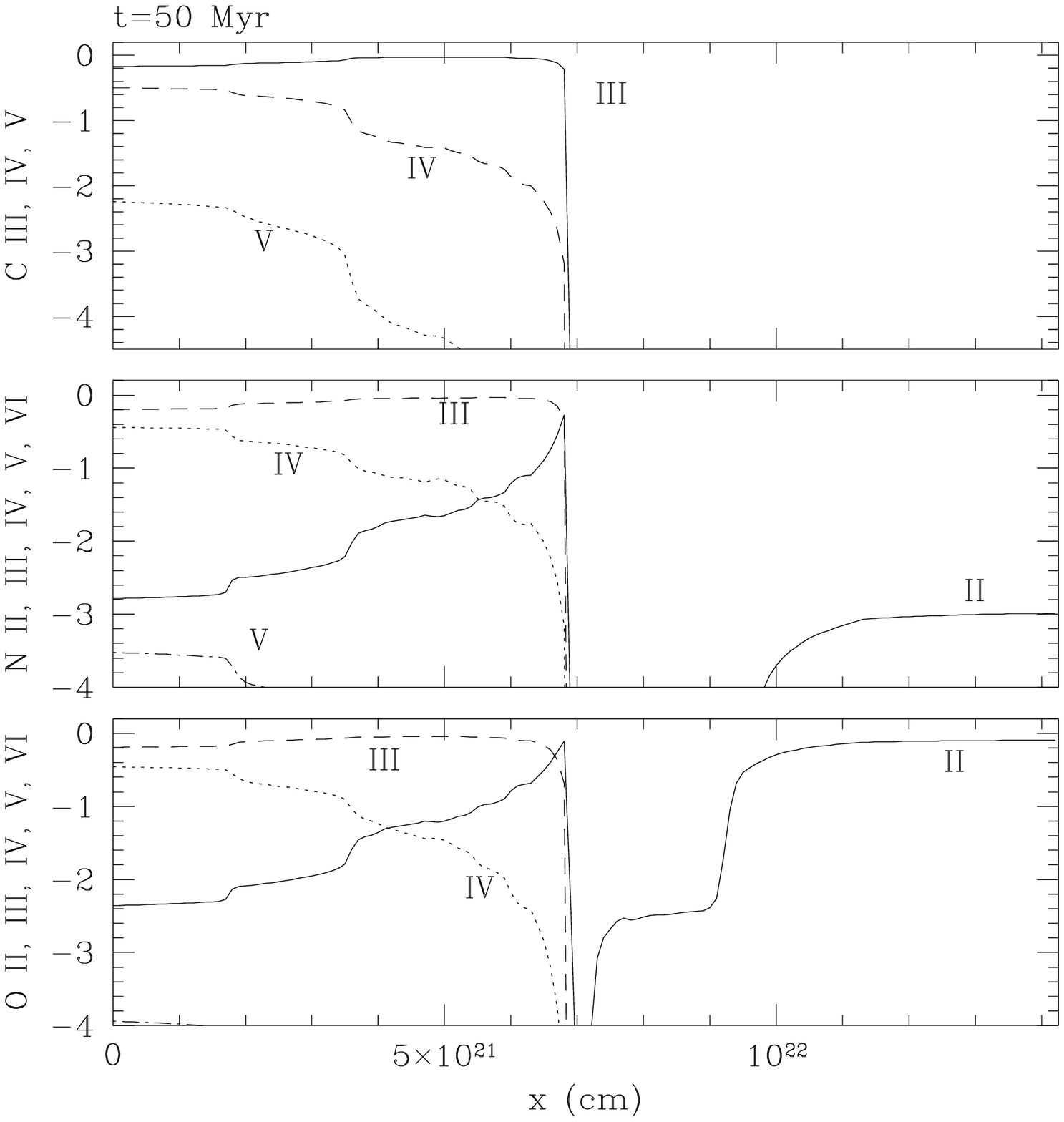}
\includegraphics[width=0.5\textwidth]{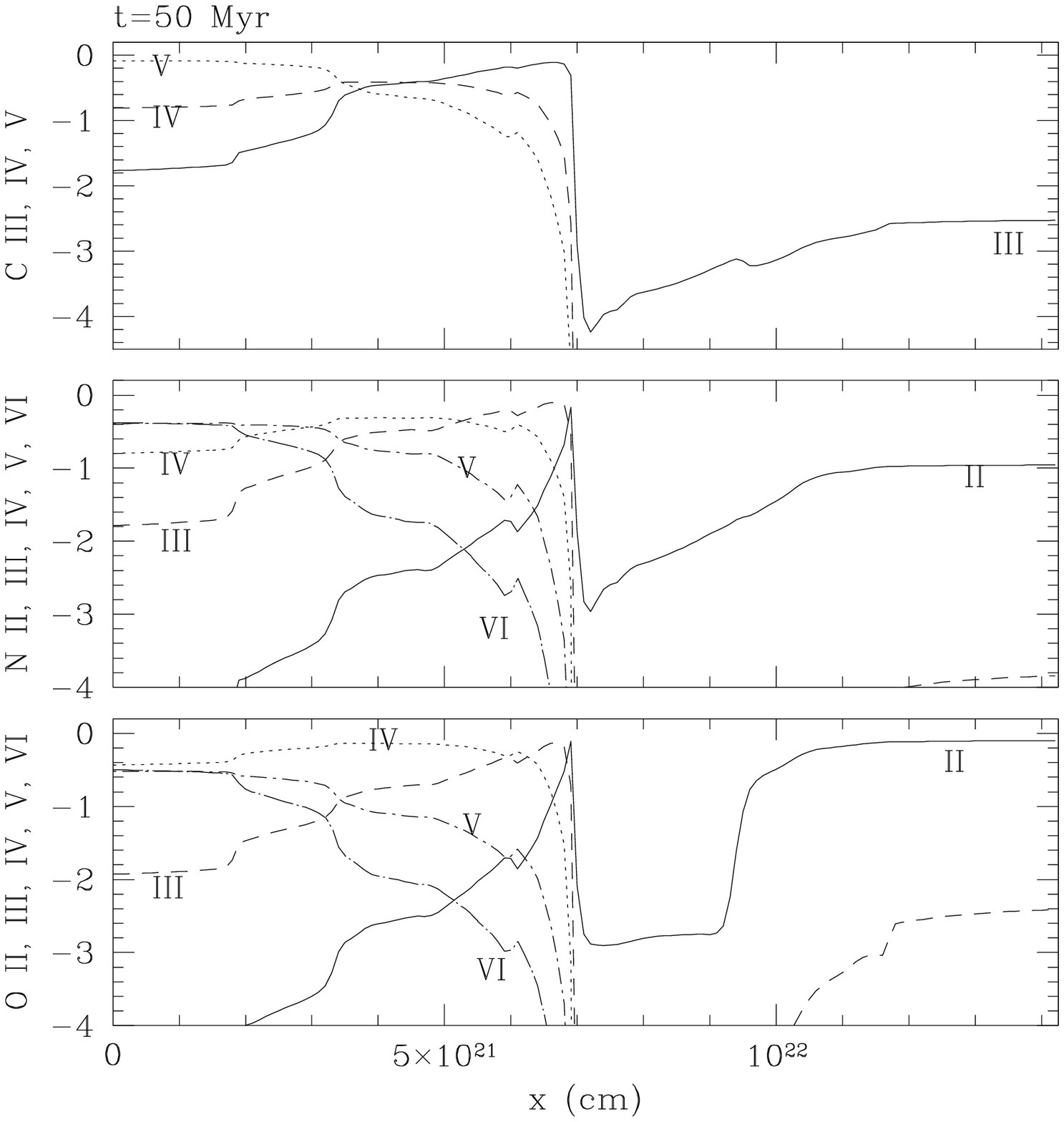}
\vspace{-1cm}
\caption{OBSERVATIONAL DIAGNOSTICS I: IONIZATION STRUCTURE OF METALS.
        C, N, and O ionic fractions along symmetry
        axis at $t = 50\rm\,Myr$. (a) (left)  STELLAR CASE; (b) (right) 
	QUASAR CASE}
\end{figure}
\begin{figure}
\vspace{-3.5cm}
\hspace{-1.2cm}
\includegraphics[width=0.6\textwidth]{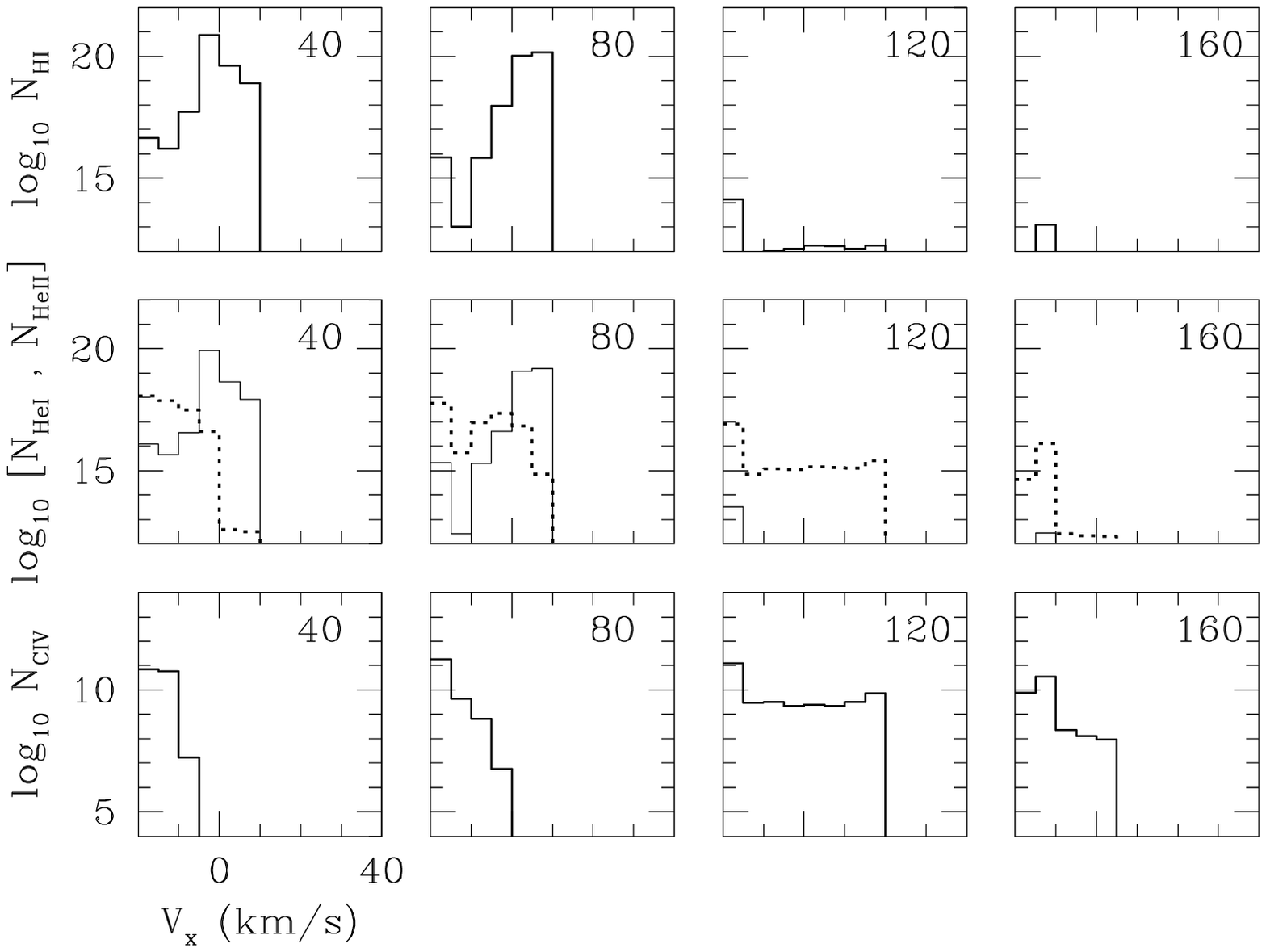}
\hspace{-1.5cm}
\includegraphics[width=0.6\textwidth]{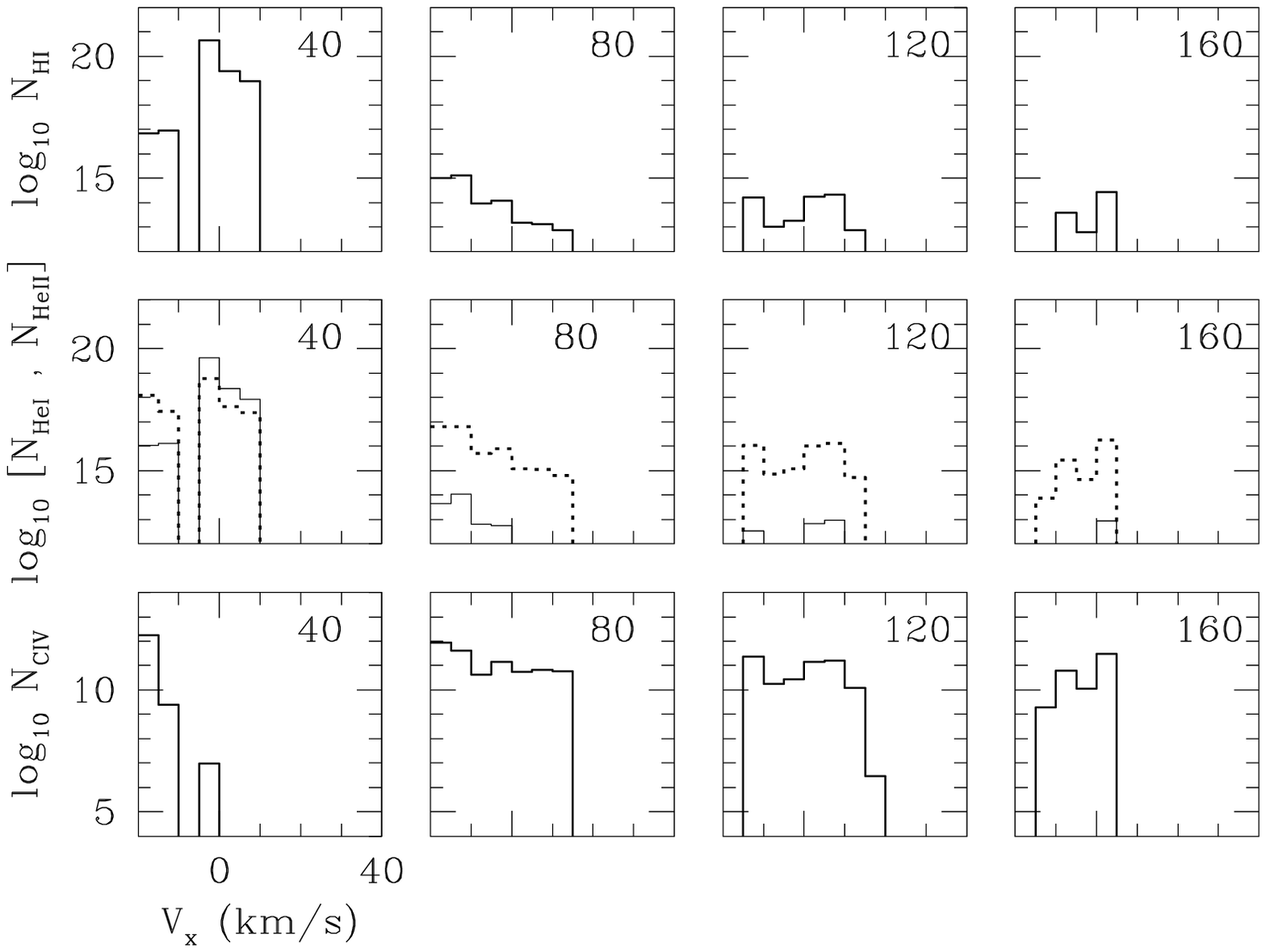}
\vspace{-2.5cm}
\caption{OBSERVATIONAL DIAGNOSTICS II. ABSORPTION LINES. 
        Minihalo column densities ($\rm cm^{-2}$) along symmetry axis at
        different velocities. (Top) H~I; (Middle) He~I (solid) and
        He~II (dotted); (Bottom) C~IV. Each box labelled with time 
        (in Myrs) since source turn-on. (a) (left)  STELLAR CASE; (b) (right) 
	QUASAR CASE.}
\end{figure}

This work was supported by grants NASA NAG5-2785, NAG5-7363, and
NAG5-7821, NSF ASC-9504046, and Texas ARP 3658-0624-1999, and a 
1997 CONACyT National Chair of Excellence at UNAM for PRS.


\end{document}